\pdfoutput=1
\documentclass{article}

\usepackage{arxiv}

\usepackage[utf8]{inputenc} 
\usepackage[T1]{fontenc}    
\usepackage{hyperref}       
\usepackage{url}            
\usepackage{booktabs}       
\usepackage{amsfonts}       
\usepackage{nicefrac}       
\usepackage{microtype}      
\usepackage{lipsum}		
\usepackage{graphicx}
\usepackage[numbers]{natbib}
\usepackage{doi}

\usepackage{xcolor}
\usepackage{caption}
\usepackage{subcaption}     
\usepackage{amsmath}
\usepackage{lscape}
\usepackage{enumitem}
\usepackage{multirow}
\usepackage{soul}
\usepackage[normalem]{ulem}

\title{Exploring the limitations of blood pressure estimation using the photoplethysmography signal}

\author{Felipe M.~Dias \\
	Heart Institute (InCor)\\
	Clinics Hospital University of \\
 Sao Paulo Medical School\\
	Sao Paulo - SP - Brazil \\
	\texttt{f.dias@hc.fm.usp.br} \\
    \And
    Diego A.C.~Cardenas \\
	Heart Institute (InCor)\\
	Clinics Hospital University of \\
 Sao Paulo Medical School\\
	Sao Paulo - SP - Brazil \\
	\texttt{diego.cardenas@hc.fm.usp.br} \\
    \And
    Marcelo A.F.~Toledo \\
	Heart Institute (InCor)\\
	Clinics Hospital University of \\
 Sao Paulo Medical School\\
	Sao Paulo - SP - Brazil \\
	\texttt{marcelo.arruda@hc.fm.usp.br} \\
 \And
    Filipe A.C.~Oliveira \\
	Heart Institute (InCor)\\
	Clinics Hospital University of \\
 Sao Paulo Medical School\\
	Sao Paulo - SP - Brazil \\
	\texttt{filipe.acoliveira@hc.fm.usp.br} \\
    \And
    Estela ~Ribeiro \\
	Heart Institute (InCor)\\
	Clinics Hospital University of \\
 Sao Paulo Medical School\\
	Sao Paulo - SP - Brazil \\
	\texttt{estela.ribeiro@hc.fm.usp.br} \\
 \And
	Jose E.~Krieger \\
	Heart Institute (InCor)\\
	Clinics Hospital University of \\
 Sao Paulo Medical School\\
	Sao Paulo - SP - Brazil \\
	\texttt{j.krieger@hc.fm.usp.br} \\
  \And
	Marco A.~Gutierrez \\
	Heart Institute (InCor)\\
	Clinics Hospital University of \\
 Sao Paulo Medical School\\
	Sao Paulo - SP - Brazil \\
	\texttt{marco.gutierrez@incor.usp.br}
}

\date{}

\renewcommand{\shorttitle}{IABP - PPG - BLOOD PRESSURE}

\hypersetup{
pdftitle={A template for the arxiv style},
pdfsubject={q-bio.NC, q-bio.QM},
pdfauthor={David S.~Hippocampus, Elias D.~Striatum},
pdfkeywords={First keyword, Second keyword, More},
}

\begin{document}
\maketitle

\begin{abstract}

Hypertension, a leading contributor to cardiovascular morbidity, underscores the need for accurate and continuous blood pressure (BP) monitoring. While traditional cuff-based methods offer periodic measurements, they fall short of providing real-time BP monitoring, driving the demand for innovative, non-invasive solutions. Photoplethysmography (PPG) presents a promising approach for such continuous BP monitoring. However, the precision of BP estimates derived from PPG signals have been subjects of ongoing debate, necessitating comprehensive evaluation of their effectiveness and constraints.

In our study, we developed a calibration-based Siamese ResNet model for BP estimation, utilizing a signal input paired with a reference BP reading.
We compared the use of normalized PPG (N-PPG) against the normalized Invasive Arterial Blood Pressure (N-IABP) signals as input.
The N-IABP signals does not directly present systolic and diastolic values but theoretically provides a more accurate BP measure than PPG signals since it is a direct pressure sensor inside the body. This foundational assumption posits that if BP estimation from N-IABP signals poses challenges, predicting BP from PPG signals might be even more challenging. Hence, our strategy potentially establishes a critical benchmark for PPG performance, realistically calibrating expectations for PPG's BP estimation capabilities.
Nonetheless, we compared the performance of our models using three signal filtering conditions: i) no filtering; ii) band-pass filtering between 0.5 Hz and 10 Hz; and iii) band-pass filtering between 0.5 Hz and 3.5 Hz to evaluate the impact of filtering on the results.

We evaluated our method using the Association for the Advancement of Medical Instrumentation (AAMI) and the British Hypertension Society (BHS) standards using the VitalDB dataset.
Both N-IABP and N-PPG for all filtering configurations surpassed the baseline predictive model (that just repeated the calibration value as prediction), evidencing their ability to predict signal-to-BP correlations beyond mere calibration mimicry.
The N-IABP signals meet with AAMI standards for both Systolic Blood Pressure (SBP) and Diastolic Blood Pressure (DBP), with raw signals notably outperforming their filtered counterparts. 
Specifically, unfiltered N-IABP signals obtained an 'A' grade in BHS assessments, whereas filtering, especially within a constrained 0.5 Hz -- 3.5 Hz band, significantly reduced performance. 
In contrast, N-PPG signals exhibited minimal performance deviation between unfiltered and 0.5 Hz -- 10 Hz filtered states but faced a marked decline under stringent filtering conditions  (0.5 Hz -- 3.5 Hz).

Our findings highlight the potential and limitations of employing PPG for BP estimation, showing that these signals contain information correlated to BP but may not be sufficient for predicting it accurately. The comparison with the results from the N-IABP signal offers a realistic benchmark for future advancements in the field using PPG signals.

\end{abstract}


\section{Introduction}
\label{Introduction}


Hypertension is a major factor contributing to cardiovascular mortality, significantly increasing the risks of heart failure, stroke, and myocardial infarction \cite{CHOCKALINGAM2007}. Affecting over 1 billion adults globally, it's a prevalent health concern. Effective hypertensive treatment encompasses not only medication but also lifestyle modifications such as dietary adjustments and increased physical activity \cite{carey2022treatment}. An essential component of managing hypertension is achieving and maintaining a target Blood Pressure (BP) level, requiring regular and precise monitoring of BP \cite{carey2022treatment,  park2019ideal}. This underscores the importance of accurate and consistent BP monitoring for hypertensive patients.

Traditional BP monitoring predominantly uses sphygmomanometers with an inflatable cuff. The manual technique involves healthcare professionals using a stethoscope to detect Korotkoff sounds for accurate systolic and diastolic pressure readings \cite{ burton1967criterion, o1990british}. Alternatively, automated oscillometric sphygmomanometers measure BP by sensing arterial wall oscillations, though they may lack precision under certain conditions such as patient movement or irregular heart rhythms \cite{jones2003measuring}. 
Both methods, however, are limited by the discomfort caused by cuff inflation and their unsuitability for continuous monitoring, potentially missing vital BP fluctuations essential in hypertension management \cite{parati2006blood}.

\begin{figure}[!h]
	\centering
    \includegraphics[width=0.50\linewidth]{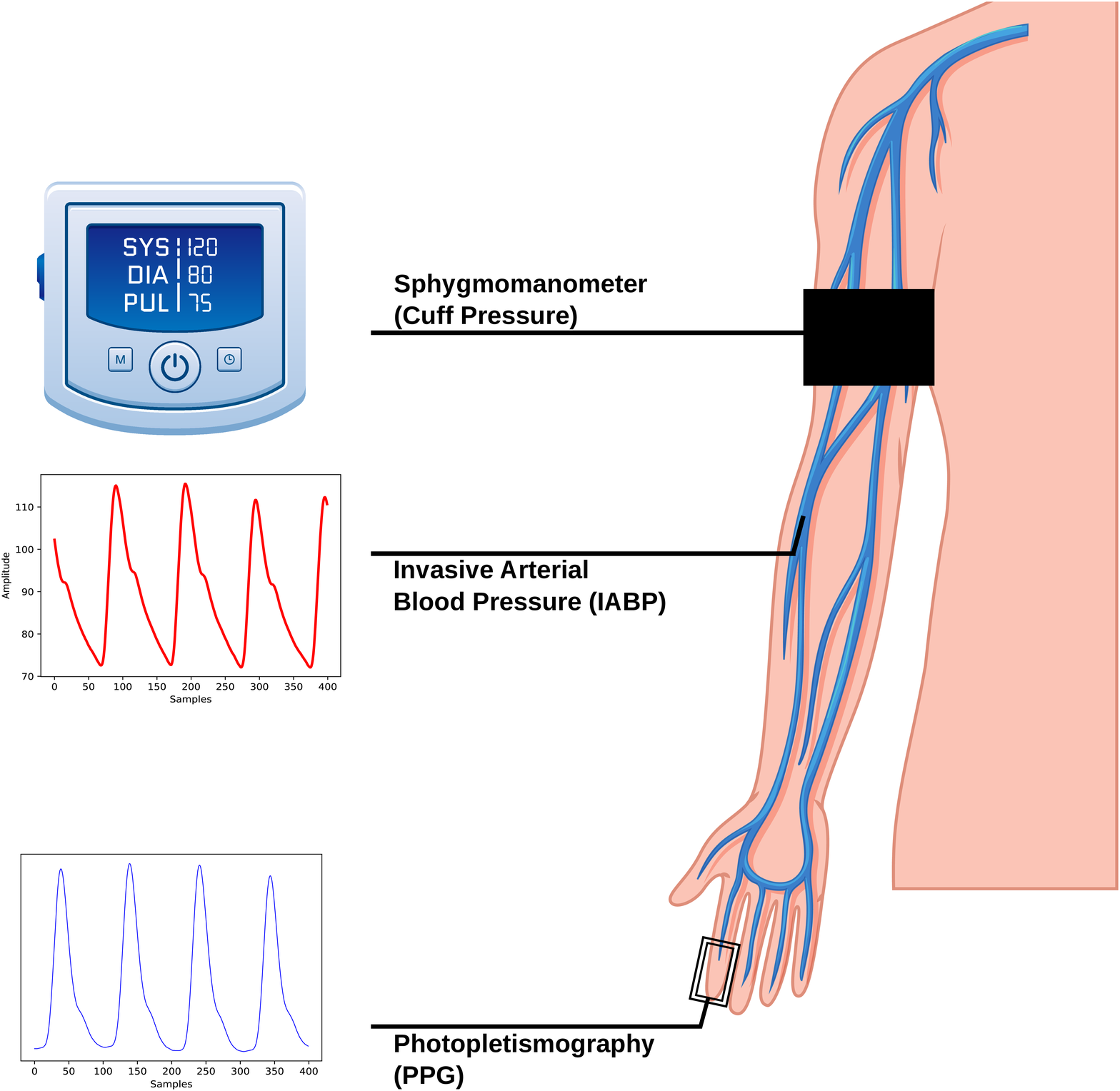}
	\caption{Illustration of different Blood Pressure Monitoring Methods.}
	\label{fig:pressao_medidas}
\end{figure}

Cuff-based methods provide instant BP measurements, capturing systolic blood pressure (SBP) and diastolic blood pressure (DBP). 
Nevertheless, arterial blood pressure (ABP) is a dynamic parameter that continually varies.
In this context, Invasive Arterial Blood Pressure (IABP) monitoring offers a more comprehensive view of these fluctuations, enabling continuous monitoring and direct accurate estimations of SBP and DBP \cite{schroeder2010cardiovascular}. 
Despite its detailed cardiovascular insights, this invasive technique requires skilled professionals for catheter insertion and management and poses risks like bleeding, infection, arterial damage, and thrombosis \cite{scheer2002clinical}. Figure \ref{fig:pressao_medidas} depicts aforementioned techniques for blood pressure monitoring.

Recent research in non-invasive and continuous BP monitoring have increasingly relied on Photoplethysmography (PPG) signals, which are optical measurements indicative of blood volume changes. Within the methodologies employed for estimating BP using PPG, two primary approaches stand out: the Pulse Wave Velocity (PWV)-based approach and the Pulse Wave Analysis (PWA) method \cite{el2020review, mukkamala2022cuffless}.

The PWV-based method estimates BP by measuring the velocity of the pulse wave traveling through the blood vessels \cite{deb2007cuff}. It employs techniques like Pulse Transit Time (PTT) -- the time taken for the pulse wave to travel between two arterial sites -- or Pulse Arrival Time (PAT) -- the time from the cardiac cycle's start to the arrival of the pulse wave at a peripheral site. These measurements help estimate PWV, which is then related to BP using theoretical equations such as the Bramwell–Hill and Moens–Korteweg equation \cite{jc1922velocity, tijsseling2012isebree}. Calibration is essential in these methods, necessitating the concurrent collection of PPG signals and validated BP measurements to properly estimate the parameters of the BP prediction equation.
Nonetheless, the application of PTT and PAT for BP estimation presents challenges, particularly the necessity for PPG measurements at various locations (for PTT) or the concurrent recording of PPG and Electrocardiogram (ECG) signals (for PAT).

In contrast, the PWA is an approach where PPG waveform characteristics are extracted to predict BP, often using machine learning techniques \cite{mousavi2019blood, hasanzadeh2019blood}. 
This approach is marked by its utilization of single-site PPG measurements, significantly simplifying the data collection process compared to methods reliant on PTT and PAT. 
A variant of the PWA approach is to use deep learning for both feature extraction and regression analysis for BP prediction \cite{slapnivcar2019blood, panwar2020pp}. 
Additionally, while these models may eliminate the need for calibration, thereby broadening their applicability, some studies incorporate calibration for personalized predictions and enhanced accuracy \cite{aguet2021feature}.

It is worth noting that, despite the high correlation between PPG and IABP signals \cite{martinez2018}, the reliability of PPG in providing sufficient information for accurate BP estimation is debatable. Although several studies report promising outcomes \cite{panwar2020pp, ibtehaz2020ppg2ABP}, recent research has highlighted methodological issues such as data leakage, casting doubt on these findings \cite{dias2022machine, da2023blood, mehta2023can, weber2023intensive}.

To address the question of whether PPG signals are capable to provide enough information for the estimation of BP, our study proposes an approach that utilizes  normalized IABP (N-IABP) signals for BP estimation. This approach eliminates the explicit values of SBP and DBP, placing emphasis on analyzing the waveforms.
We hypothesize that if successful estimation proves challenging with the IABP waveform, it would likely be even more difficult with the PPG signal.
This benchmark would represent the highest achievable accuracy using PPG signals in a given dataset.
Through this approach, our study seeks to provide a realistic assessment of the capabilities of PPG signals in estimating BP, thereby establishing a clearer understanding of their  possible potential and limitations.

We propose a calibration-based pipeline for BP estimation using deep learning, leveraging either single-site PPG or IABP signals. The methodology is applied to the VitalDB dataset with signals registered from surgical patients \cite{lee2022vitaldb}. Signal preprocessing, segmentation into standardized time windows, bandpass filtering, and quality analysis constitute our approach. High-quality signals are matched with corresponding calibration signals from the same patient to serve as references for our neural network. Furthermore, the neural network, designed based on a Siamese ResNet architecture, extracts features from signal pairs and integrates them with calibration BP values to predict the BP of the target inference signal.
Additionally, we implemented different filtering bands on the signal windows and evaluated their impact on the classifier's performance to elucidate the extent to which the filters could either impede or enhance the BP estimation task.
In summary, our paper offers the following key contributions:
\begin{itemize}
\item Establishment of a benchmark for BP prediction accuracy using N-IABP waveforms, serving as a reference for assessing PPG-based methods.
\item Development of a deep learning pipeline for BP estimation from both PPG and IABP signals, using data from a large-scale surgical patient dataset.
\item Comparative analysis of PPG and IABP signal performance under identical experimental conditions, evaluating their efficacy in BP prediction.
\end{itemize}
\section{Background}



\subsection{Literature Review}

The literature on blood pressure estimation using single-site PPG divides methodologies into two distinct categories: those that require calibration (calibration-based approaches) and those that do not (calibration-free approaches).

Calibration-based approaches require a preliminary step that involves the user first obtaining a reference PPG signal, along with corresponding systolic and diastolic blood pressure values, using a validated equipment. The need for this initial calibration step, which requires specific equipment and additional effort, can make these approaches less user-friendly and more cumbersome in practical applications.
On the other hand, calibration-free approaches simplify the process significantly. In this method, the user is only required to acquire a PPG signal. The model predicts BP values directly from this signals, without the need for any prior  reference data. This approach is more straightforward and user-friendly, as it eliminates the extra step of obtaining calibrated BP readings.
However, the absence of a BP reference point introduces its own set of challenges, making these approaches inherently more complex to develop and potentially less accurate in practice.

Calibration-free approaches predominate in the literature on BP estimation using PPG signals. 
For example, 
Slapni{\v{c}}ar et al. \cite{slapnivcar2019blood} employed a spectro-temporal deep neural network, integrating residual connections. Using the MIMIC III dataset, their model obtained MAE of 9.43 mmHg for systolic BP and 6.88 mmHg for diastolic BP. 
Panwar et al. \cite{panwar2020pp} introduced ``PP-Net'', a deep learning model designed for simultaneous estimation of SBP, DBP, and heart rate (HR) from the PPG signal. The model's performance was validated on the MIMIC-II-UCI dataset \cite{kachuee2015cuff, kachuee2016cuffless, cuff}, achieving normalized MAE of 0.09 for DBP, 0.04 for SBP, and 0.046 for HR estimation. 
In addition, Wang et al. \cite{wang2024imsf} proposed the IMSF-Net, an improved multi-scale information fusion network specifically tailored for PPG-based BP estimation. This network was evaluated using the MIMIC-II dataset achieving MAE of 2.07 $\pm$ 2.32 mmHg for SBP and 1.21 $\pm$ 1.51 mmHg for DBP.
Lastly, El-Hajj et al. \cite{el2021cuffless} focused on extracting features from PPG waveforms and employing advanced deep learning techniques, including bidirectional recurrent layers and an attention layer. The model achieved a MAE of 4.51 $\pm$ 7.81 mmHg for SBP and 2.6 $\pm$ 4.41 mmHg for DBP.

Although more practical, calibration-free approaches do not generally present good estimation results.
A critical examination of these works, mainly the ones using the MIMIC-II-UCI dataset \cite{kachuee2015cuff, kachuee2016cuffless, cuff}, uncovers significant flaws. This dataset is widely used in this field because it is already preprocessed and it's easy to use compared to the original MIMIC II dataset \cite{saeed2011multiparameter} but it lacks patient identification information, leading to a common data leakage validation issue where patient data appear in both training and testing sets \cite{dias2022machine, da2023blood}.  


In this context, using a calibration-based approach seems more promising and realistic. 
As an example of this approach, Aguet et al.\cite{aguet2021feature} proposed a siamese convolutional neural network to tackle the BP estimation challenge using the VitalDB dataset \cite{lee2022vitaldb}. 
This model leverages mean PPG cycles and their derivatives, coupled with calibration pressures. 
This work ensured that data from the same patient is not employed across train and test sets, i.e., no data leakage. The model achieved an accuracy of -0.24$\pm$10.27 mmHg for systolic and -0.50$\pm$6.52 mmHg for diastolic blood pressure, indicating a promising direction for future research.

Due to the large number of works in the field that have methodological errors, this field of study has received some important critiques recently. 
Weber et al. \cite{weber2023intensive} address generalization issues in BP estimation across different datasets, suggesting limitations in current methodologies when applied to diverse real-world scenarios. Furthermore, Dias et al. \cite{dias2022machine} and Costa et al. \cite{da2023blood} explored the inherent limitations of the MIMIC II-UCI dataset, particularly highlighting how the absence of patient identification compromises the integrity of the data splitting process. Their investigation centered on the analysis of various validation schemes prone to data leakage, which can lead to the artificial inflation of model performance metrics. This critical examination underscores the potential for misleading conclusions regarding the effectiveness of PPG signals in BP prediction, spotlighting the necessity for stringent validation practices to ensure the accuracy and reliability of research findings in this domain.
Also, Mehta et al. \cite{mehta2023can} provide a thorough investigation into the intrinsic capabilities of PPG signals in predicting BP. Their work present common flaws in the literature such as excessive data contraction and unrealistic preprocessing, concluding that BP estimation using PPG signals might be an ill-posed problem with significant challenges yet to overcome.
These papers, although scarce in the literature, raise important fundamental doubts about whether PPG signals inherently contain sufficient information for accurate BP prediction or if this represents an insurmountable challenge.

\subsection{Theoretical Concepts}

\subsubsection{Photoplethismography (PPG)}


PPG is an optical measurement technique that detects blood volume changes \cite{rolfe1979photoelectric}.
This technique has found widespread clinical application, particularly in devices like pulse oximeters and smartwatches \cite{CHARLTON2022401}. 
It requires a cheap hardware setup: a light source to illuminate the tissue and a photodetector to measure the variations in light intensity that are indicative of changes in blood volume \cite{allen2007photoplethysmography}. 

The PPG signal comprises two primary components: the ``AC'' component, which fluctuates with each heartbeat, and the ``DC'' component, representing the baseline blood volume and tissue properties. Additionally, the PPG waveform illustrates the heart's rhythmic activity through two phases: the anacrotic phase, marking the wave's ascent during systole (heart contraction), and the catacrotic phase, its descent during diastole (heart relaxation). The systolic peak, at the anacrotic phase's conclusion, signals the maximum blood ejection, while the catacrotic phase ends with the diastolic peak, indicating heart filling. The dicrotic notch, a distinct dip following the systolic peak, signifies the aortic valve's reopening as the heart prepares for the next cycle \cite{park2022photoplethysmogram}.

\begin{figure}[htbp]
    \centering
    \begin{subfigure}[t]{0.45\textwidth}
        \includegraphics[width=\textwidth]{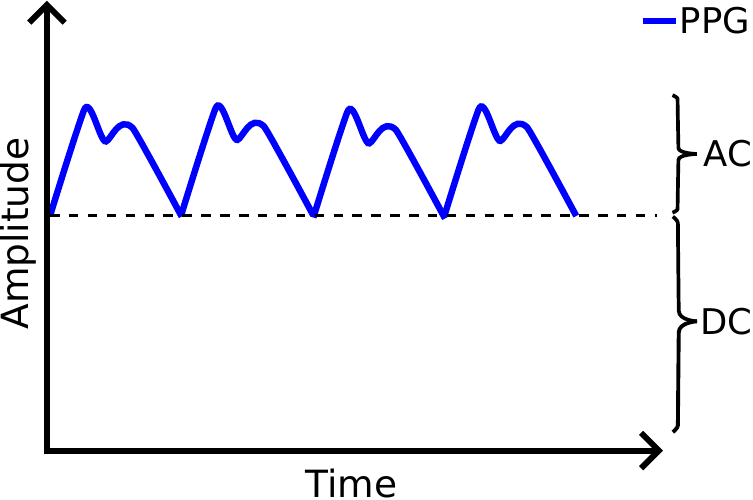} 
        \caption{Oscillating (AC) and constant (DC) components of the PPG signal.}
        \label{fig:ppg_dc_ac}
    \end{subfigure}
    \hfill
    \begin{subfigure}[t]{0.45\textwidth}
        \includegraphics[width=0.65\textwidth]{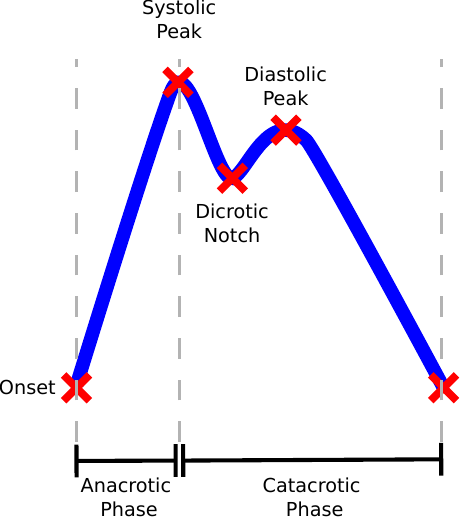}
        \caption{PPG waveform with key features: onset, systolic peak, dicrotic notch, and diastolic peak, showing cardiac cycle phases}
        \label{fig:ppg_waveform}
    \end{subfigure}
    \caption{Comprehensive analysis of PPG signal components and waveform characteristics.}
    \label{fig:ppg}
\end{figure}

Figure \ref{fig:ppg} shows both the PPG AC and DC components along with the PPG waveform main characteristics.

PPG signals can be obtained from various body parts, each exhibiting unique characteristics influenced by the specific region \cite{hartmann2019quantitative}. 
Usually, these signals are predominantly captured from the finger, a method widely utilized in hospital oximeters.
Recent advancements in wearable technology have expanded the application of PPG sensors to include smartbands and smartwatches. 
These devices typically capture PPG signals from the wrist.

\subsubsection{Invasive Arterial Blood Pressure (IABP)}


BP measures how strongly blood pushes against the walls of arteries and is given in mmHg (millimeters of mercury). It varies with the heart's rhythm, increasing as the heart pumps blood out (systolic pressure) and decreasing as the heart refills (diastolic pressure). Systolic pressure is the peak pressure when the heart contracts, specifically from the left ventricle, which sends blood throughout the body. Diastolic pressure is the lower pressure when the heart rests between beats and refills with blood \cite{AHA2021}.

IABP monitoring is considered the gold standard regarding BP measurement \cite{meidert2018techniques, goodman2023measuring}.
It is a crucial technique in hemodynamic monitoring, particularly vital in the care of critically ill patients in intensive care units and during high-risk surgical procedures \cite{lakhal2012noninvasive}. 
The great advantage of IABP technique is that it enables direct and real-time BP measurement as opposed to other approaches (auscultatory) \cite{WARD2007122}. 

This monitoring is typically achieved through the cannulation of a peripheral artery, such as the radial, brachial, femoral, or dorsalis pedis artery. 
The process involves the transmission of pressure exerted by cardiac contractions through a fluid-filled catheter to a transducer. This transducer then converts the mechanical motion into electrical signals, displaying both a beat-to-beat arterial waveform and numerical pressures on a monitor \cite{WARD2007122}.

However, invasive monitoring has its drawbacks. It's contraindicated in cases like infection at the insertion site and compromised collateral circulations. Complications can include infections, damage to nearby nerves, and thrombosis. These highlight the critical need for precision and vigilance during medical interventions. \cite{WARD2007122}.

\begin{figure}[!h]
	\centering
    \includegraphics[width=0.65\linewidth]{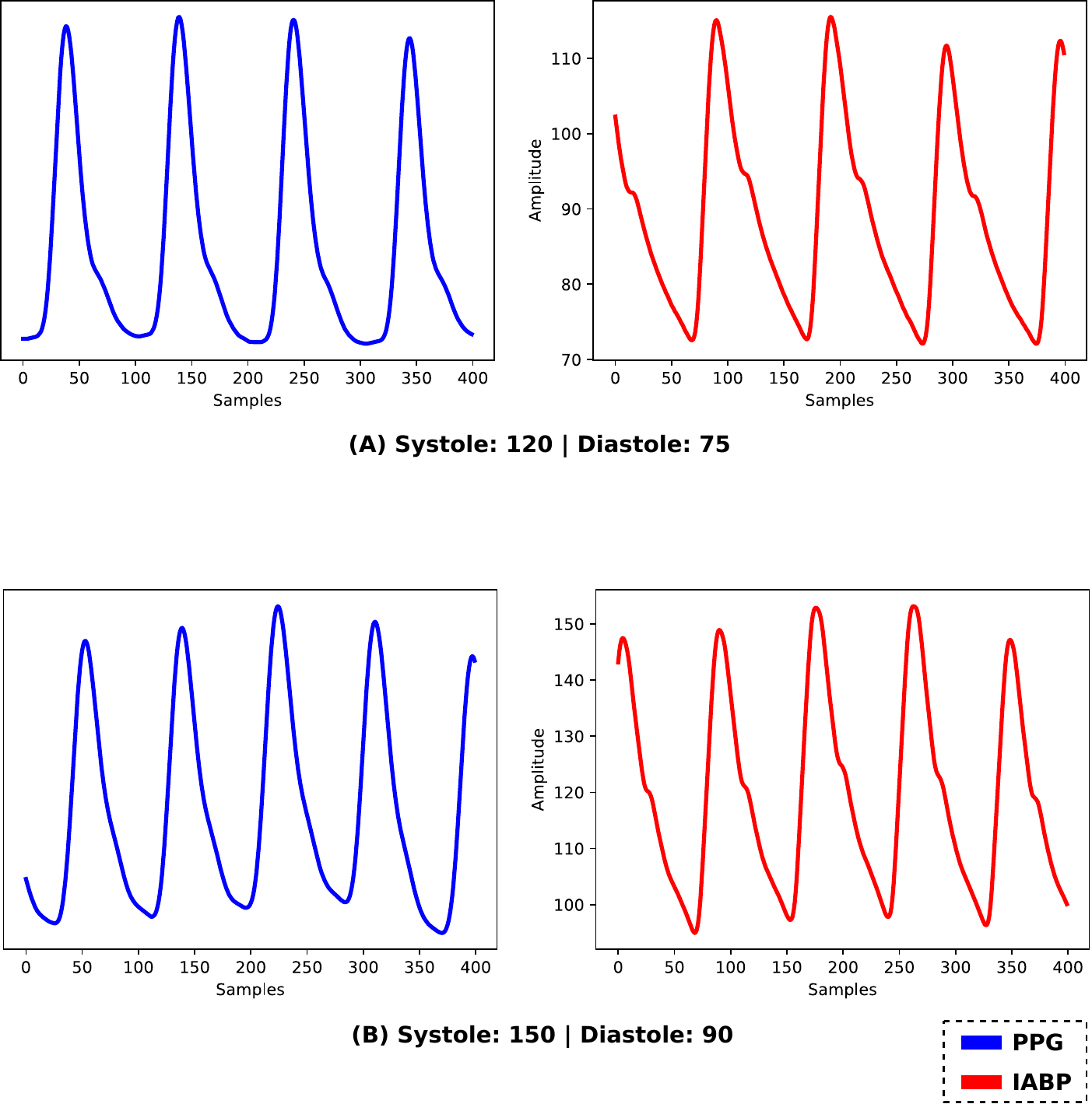}
	\caption{Pairs of PPG (blue) and IABP (red) waveform samples extracted from the same patient in the VitalDB dataset for two distinct instant in time. Systolic/Diastolic values for both selected windows are labeled as (A) 120/75 mmHg, and (B) 150/90 mmHg. 
    }
	\label{fig:abp_wave}
\end{figure}

The relation between PPG and IABP signal can be seem due to their morphological resemblance.
As shown in Figure \ref{fig:abp_wave}, which presents two synchronous pairs of PPG and IABP signals from the same patient within the VitalDB dataset \cite{lee2022vitaldb}, a notable morphological resemblance between the PPG and IABP waveforms is observed. 
This observation is corroborated by the findings of Martinez et al. \cite{martinez2018}, who provided evidence supporting the assertion that PPG signals contain information analogous to that of the IABP signal.

However, it is crucial to note the significant disparity in systolic and diastolic pressures between the PPG/IABP signal pairs (120/75 mmHg in the top pair and 150/90 mmHg in the bottom pair). This substantial variation in BP measurements does not manifest as prominently in the PPG signals, which appear remarkably similar. Given the challenge of distinguishing between different BP labels with such similar input signals, the task of accurately correlating PPG signals with specific BP measurements is evidently complex.

\section{Methodology}
\label{Methodology}

{\subsection{Study Design}

Our proposed methodology uses a calibration approach to predict BP.  We require three inputs:


\begin{enumerate}
    \item \textbf{Calibration Signal:} Can be either the normalized PPG (N-PPG) signal or the normalized IABP (N-IABP) signal acquired during the calibration process;
    \item \textbf{Systolic and Diastolic values:} Instantaneous value of SBP and DBP acquired during the calibration process;
    \item \textbf{Inference Signal:} The N-PPG or the N-IABP signal intended to be inferred.
    
\end{enumerate}

As aforementioned, our hypothesis is that even though the N-IABP does not contain the explicit values of the systolic and diastolic pressure, its morphology contain enough information to provide a more direct measure of BP, resulting in better results. Thus, our aim is to compare the performance of N-PPG and N-IABP to measure BP.

Our study is based on a publicly available dataset that will be described in the next section. To ensure data quality, we preprocessed the signals by filtering them, checking their quality, and removing noisy segments. Only high-quality windows from both IABP and PPG signals were retained for consistency.

Moreover, since we use calibration, we established a ``pair montage'' approach. This means we created pairs of signals for training: one being the inference signal  and the other being the calibration signal (with systolic and diastolic pressure). These pairs were selected based on specific criteria. After this, all signals were normalized.

Finally, we implemented two Siamese ResNet model. One utilizing N-IABP signals, while the other utilized N-PPG signals. Both models aimed to estimate systolic and diastolic BP. Additionally, we compared the results from both models to understand how well normalized IABP and PPG signals perform in predicting blood pressure.

A summary of our proposed methodology is displayed in Figure \ref{fig: methodology} and in the next sections we'll explore in detail each step of the process.

\begin{figure}[!h]
	\centering
    \includegraphics[width=1\linewidth]{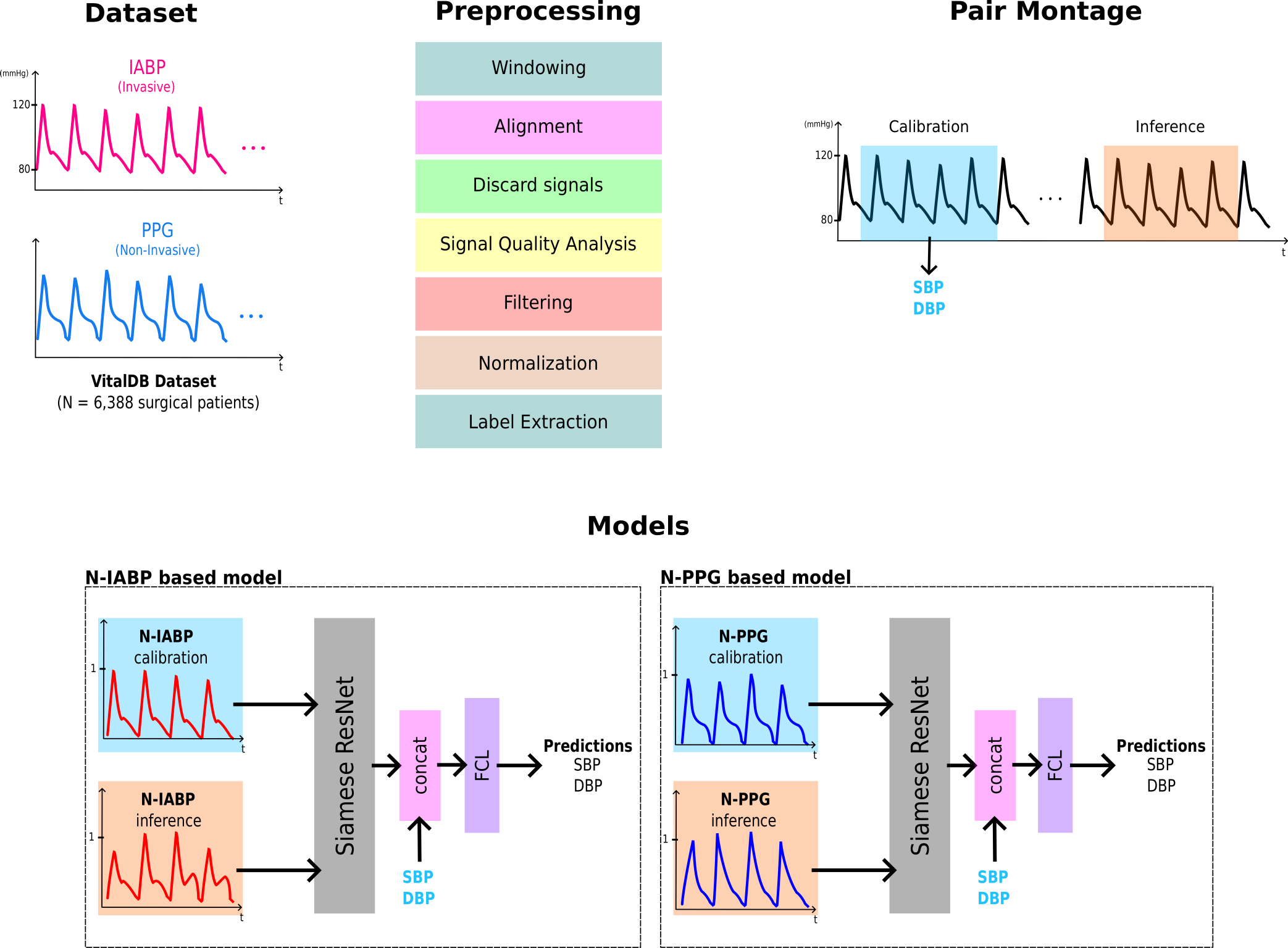}
	\caption{The proposed methodology for BP estimation. }
	\label{fig: methodology}
\end{figure}

\subsection{Dataset}
We used a public available dataset called VitalDB \cite{lee2022vitaldb}.
This dataset contains 6,388 patients that are going through surgery.
The patients get several signals continuous recorded during the procedure (e.g., PPG and IABP).
But not all patients get all signals recorded.
For our purposes, we only selected patients that have PPG and IABP, resulting in 3,338 patients in total.

\subsection{Preprocessing}
We employed some preprocessing steps
that are described in the following and illustrated on Figure \ref{fig: preprocessing}.

\begin{figure}[!h]
	\centering
    \includegraphics[width=1\linewidth]{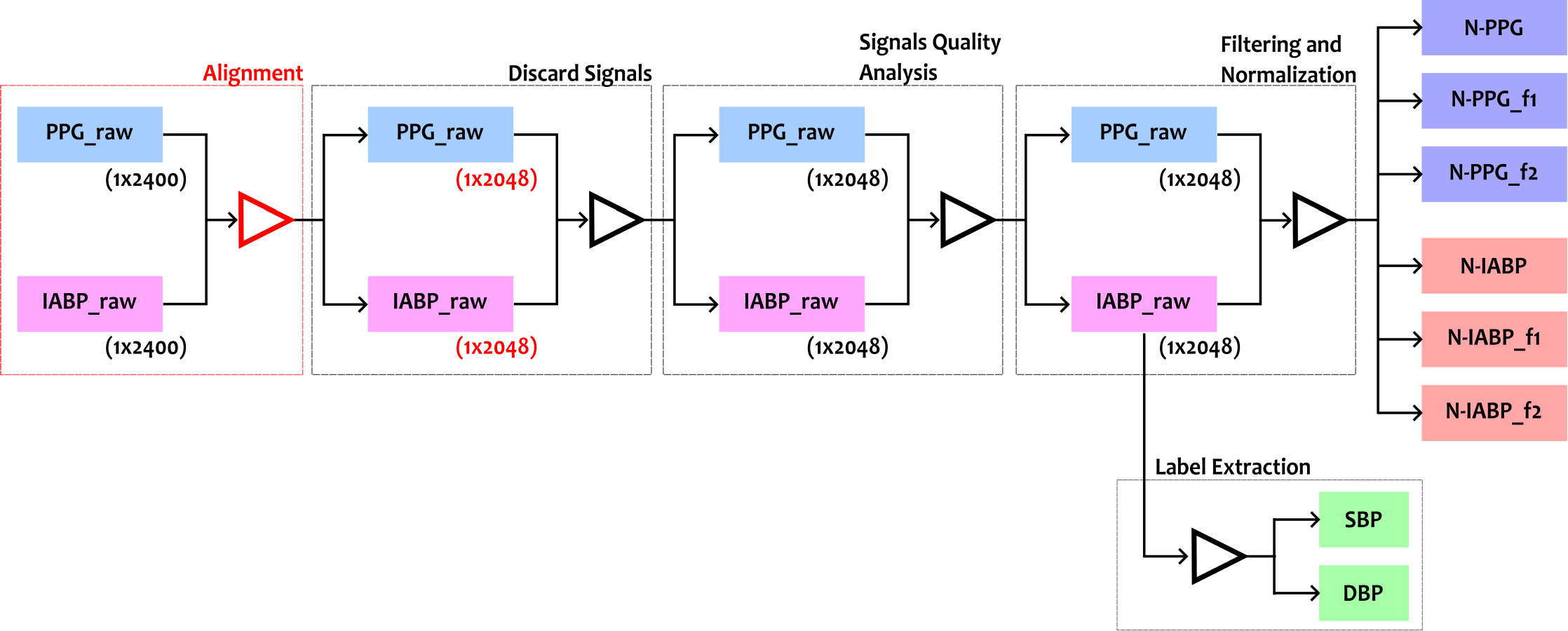}
	\caption{General preprocessing steps.}
	\label{fig: preprocessing}
\end{figure}

\subsubsection{Windowing}\label{sub:window}
The dataset comprises recordings spanning several hours for each patient. Hence, it necessitates a strategic approach for effectively conditioning this data. We chose to split the recordings from each patient into segments of 24 seconds without any overlap between the segments. Since the sampling rate of the dataset is 100 Hz, our windows comprise 2400 samples, which will be the chosen input size for our model.

\subsubsection{PPG and IABP alignment}

The PPG and IABP signals were aligned in time. They inherently lag in relation to each other, meaning their peaks do not coincide in time. Believing that a more accurate comparison could be achieved if the network processed these signals with identical time alignment, we applied cross-correlation between them to identify the lag. Subsequently, we adjusted for this lag to synchronize the signals in time. The maximum lag was 176 samples. Thus, both signals were cut at the extremes, resulting in signals with 2048 samples. We believe that this synchronization would enhance the accuracy of comparing the performance of these two signals for BP estimation. 

\subsubsection{Discard signals}
\label{sub:sc}

To enhance the quality of data and mitigate the impact of noise, we implemented the following heuristic rules for discarding signals:

\begin{itemize}
    \item Discard signals if the difference in calculated heart rates between PPG and IABP exceeds 5\%.
    \item Discard signals with heart rates below 40 BPM or above 180 BPM.
    \item Discard signals if they have systolic or diastolic blood pressure values  that  are more than 2.5 standard deviations from the mean blood pressure within the window.
    \item Discard signals with systolic blood pressure below 70 mmHg or above 200 mmHg, or diastolic blood pressure below 50 mmHg or above 140 mmHg.
\end{itemize}

\subsubsection{Signal Quality Analysis}
\label{sub:sqa}

For each matched window containing PPG and IABP signals, we employed a signal quality assessment approach similar to Dias et al. \cite{dias2022machine}. 
First, the PPG signal is filtered with a 4th-order Chebyshev Type II band-pass filter [0.5 - 10 Hz]. 
Then, for both the filtered PPG and the raw IABP signal, we extracted a mean beat by detecting beat onsets and offsets, normalizing each beat, and then averaging the detected beats.
The onsets and offsets were detected using the algorithm described in Bishop et al. \cite{bishop2018multi}. If the mean Pearson correlation between the mean beat and all detected beats exceeded a threshold of 0.9, the window was classified as valid; otherwise, it was discarded. To be classified as valid, both the corresponding PPG and IABP signals had to meet this criterion. 

\subsubsection{Filtering}\label{sub:filt}

 
After discard signals, we implemented different signals filtering, aiming to elucidate the extent to which the filters could either impede or enhance the BP estimation task.
We applied signal filtering techniques to the signal windows employing the filter proposed by Liang et al. \cite{liang2018optimal}, originally designed for PPG signals, but we adapted for use with IABP signals as well. This filter is a 4th-order Chebyshev Type II band-pass filter. 
For comparison purposes, we selected two distinct filtering bands: 0.5 Hz - 10 Hz, and 0.5 Hz - 3.5 Hz. The former represents a commonly used filter band in PPG literature, while the latter corresponds to the heart rate filter band (30 BPM - 210 BPM). We compared these filtering techniques with the use of raw normalized signals.


\subsubsection{Normalization}\label{sub:norm}

To preprocess the IABP signal windows, both the filtered signals and the raw signal underwent normalization using a max-min normalization technique. This vital step aimed to remove the systolic and diastolic components -- representing peaks and troughs of the signal -- leaving only the morphological information for the network's consideration. Ensuring fair comparison between signals, we applied identical normalization to the PPG signal. The resulting normalized IABP and PPG signals are denoted as N-IABP and N-PPG, respectively.

\subsubsection{Label extraction}
\label{sub:label}

Our model was designed to estimate systolic and diastolic BP from input signals using a calibration-based approach. Consequently, for each window of the input signal (either N-PPG or N-IABP), it is necessary to obtain the corresponding label. To achieve this, for every input window, we acquire the corresponding IABP (in its raw, unnormalized form). We then identify the peaks using the algorithm described in Bishop et al. \cite{bishop2018multi} , and calculate the median of the detected peaks and valleys. These medians are used as the systolic and diastolic pressure values for that specific window.

\subsection{Pair Montage}

Our methodology requires pairs of inference and calibration windows for functioning. 
To identify the appropriate pairs, we employed two specific criteria: 

\begin{enumerate}
    \item The temporal gap between the paired signals should range from 3 minutes to 2 hours.
    \item The difference in systolic blood pressure between the pairs must not exceed 60 mmHg.
\end{enumerate}

These steps were chosen to prevent the calibration and inference signals from being too temporally proximate, which could potentially result in similar morphology and lead to overoptimistic results. The second condition aims to limit the variation in BP changes, imposing a constraint on the model's predictive range. Thus, predictions with a difference above 60 mmHg would be discarded during the testing phase.

The process of pairing is illustrated in Figure \ref{fig: pair_montage}. 
We offer the statistics regarding the pairs in Table \ref{tab:vital_stats}.

We also imposed restrictions on the selection of patients. Patients with less than 500 pairs were excluded. If a patient had more than 750 pairs, only 750 window pairs of the patient were selected randomly. Ultimately, only 1,068 of the initial patient cohort met our established criteria.

\begin{figure}[!h]
	\centering
    \includegraphics[width=0.85\linewidth]{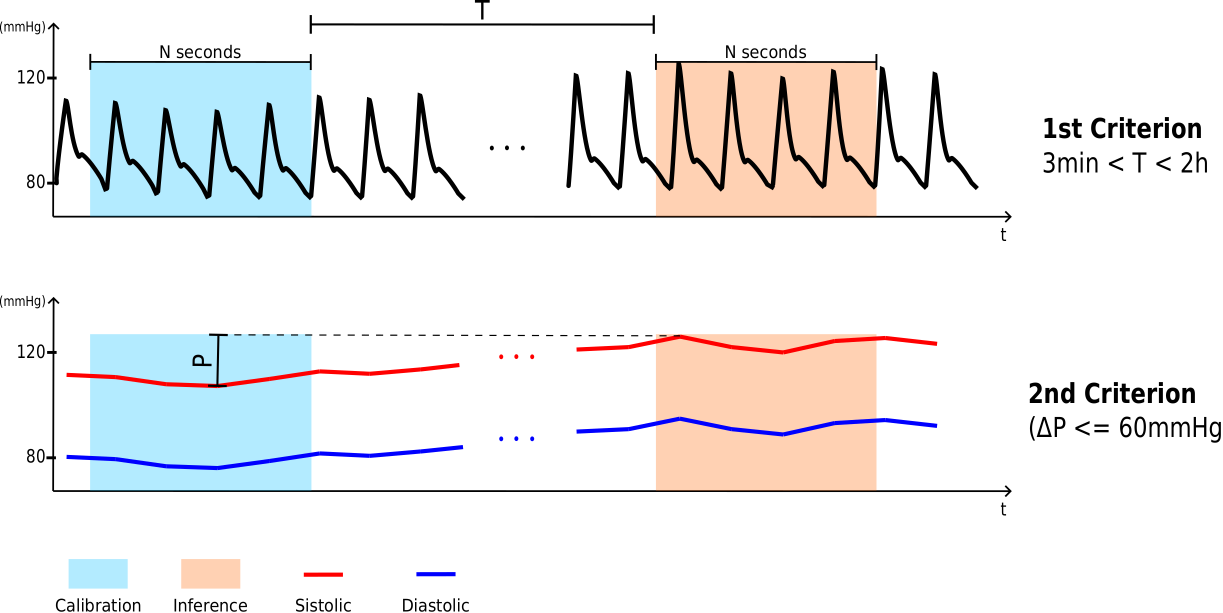}
	\caption{Pair montage selection criteria illustration.}
	\label{fig: pair_montage}
\end{figure}


In Table \ref{tab:vital_stats}, we present the statistics for SBP and DBP alongside the statistics for our calibration-inference pairs post-application of our preprocessing steps for our employed VitalDB dataset.

\begin{table}[]
\centering
\caption{\centering Statistics for SBP and DBP in the VitalDB dataset after preprocessing, including pair-montage analysis ($\Delta$DBP and $\Delta$SBP) and patient data. }
\label{tab:vital_stats}
\begin{tabular}{cccccc}
\hline
      & \textbf{DBP} & \textbf{SBP} & \textbf{} & \textbf{ $\Delta$DBP} & \textbf{ $\Delta$SBP}    \\ \hline
Mean & 64.26 & 118.94 &  & 3.42 e-8 & 6.06 e-8  \\
Standard Deviation  &  9.81 &  17.69 &  & 11.34    & 21.25   \\ \hline
\end{tabular}
\end{table}

\subsection{Proposed model}

We employed a Siamese-based framework for our methodology, illustrated in Figure \ref{fig: model}. Within this framework, we utilized a ResNet-based architecture, as described in \cite{ribeiro2020automatic}, which serves as our base network. This network accepts input sizes of 2048x1 for both calibration and inference signals.

The base network starts by processing the input through a 1D convolutional layer. After this, the output undergoes a batch normalization and is then passed through a ReLU activation function.
Following this initial processing, the data flows into a series of ResNet-like blocks. Each of these blocks operates in a specific manner:

\begin{enumerate}[label=(\alph*)]
    \item The block accepts two inputs. The first input goes through:
    \begin{itemize}
        \item A convolutional layer with $N$ filters.
        \item Batch normalization.
        \item A ReLU activation.
        \item A dropout layer, set at a 20\% rate.
    \end{itemize}
    \item The output from this sequence is then passed to another convolutional layer. This time, the convolution is characterized by a stride value, $S$.
    \item Concurrently, the second input to the block is processed with:
    \begin{itemize}
        \item A max pooling layer.
        \item A 1x1 convolutional layer to adjust the filter dimensions. This step ensures that the output matches the filter dimensions from the strided convolution mentioned earlier.
    \end{itemize}
    \item The outputs from the strided convolution and the 1x1 convolution are then summed. This sum produces the block's second output.
    \item Additionally, this sum is further refined by passing it through:
\begin{itemize}
    \item A batch normalization layer.
    \item A ReLU activation.
    \item Another dropout layer at 20\%.
    \item The result of this operation is block's first output.
\end{itemize}
   
\end{enumerate}

This entire ResNet-like block is then repeated four times within the network. For each iteration, the number of filters $N$ used are 128, 196, 256, and 320, respectively. 
The stride $S$ for the convolutional layers in all these blocks remains consistent at a value of 4.
All convolutional operations in the network consistently use a kernel size of 16, and all dropout operations maintain a dropout rate of 20\%.
Finally, after all ResNet blocks, a global average pooling layer is applied. 
The flatten features resulting from this operation are the output of the base network.

The siamese-network processes both calibration and inference signals (either N-IABP or N-PPG). Features extracted from base network are concatenated, along with calibration values for systolic and diastolic pressure. 
This concatenated information then three fully-connected layers, having 128, 64 and 2 units (systolic and diastolic outputs).
The first two layers used ReLU activation and the last layer (output layer) employed a linear activation.





\begin{figure}[!h]
	\centering
    \includegraphics[width=0.9\linewidth]{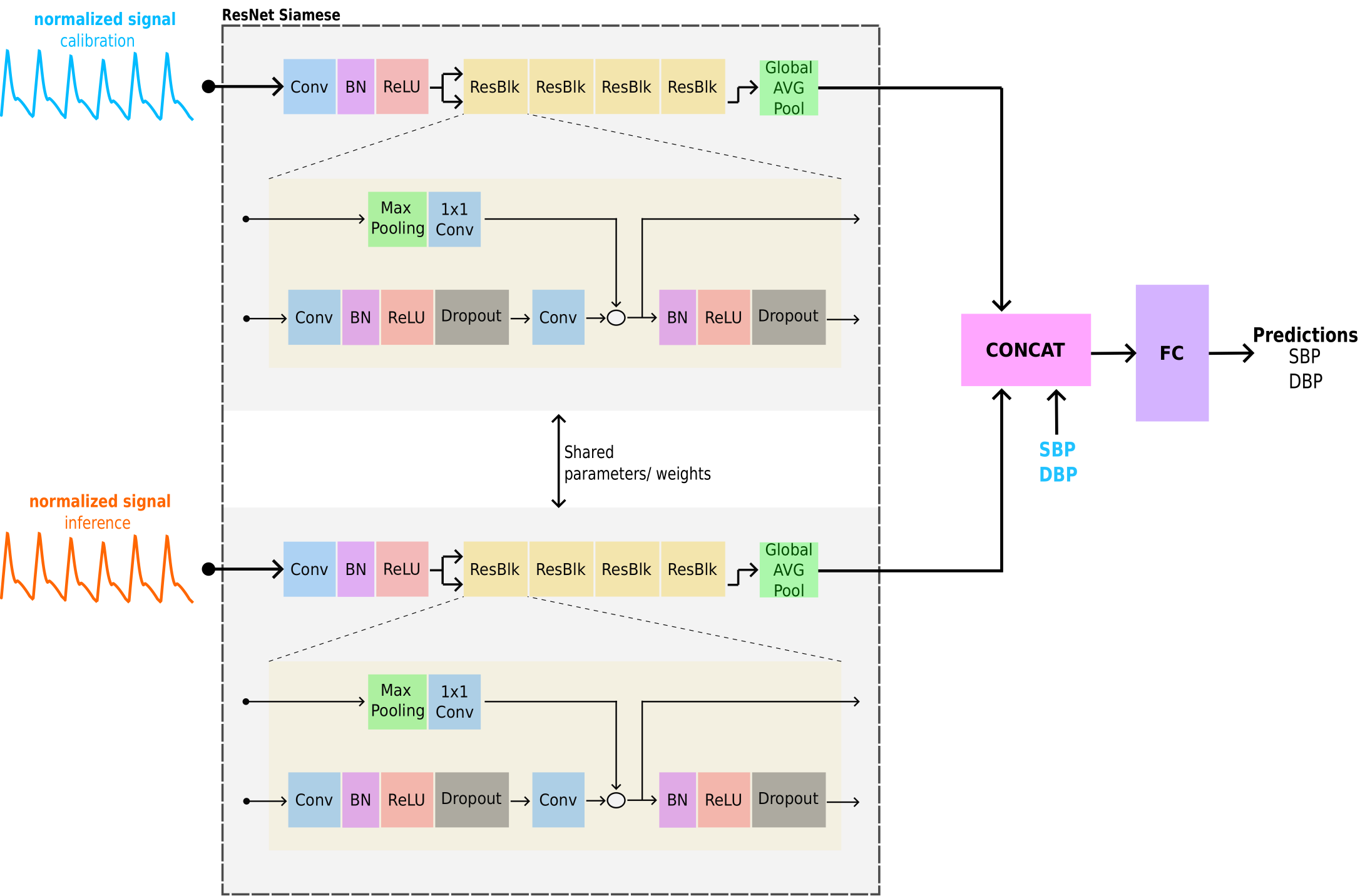}
	\caption{Proposed calibration-based Siamese Resnet network for blood pressure estimation. \\(ResBlk: ResNet block)}
	\label{fig: model}
\end{figure}

\subsection{Baseline model}

To accurately assess our proposed model's effectiveness, it is crucial to compare it against a simpler baseline model. This comparison will highlight the gains achieved through our complex model. A straightforward approach for calibration-based models would involve predicting blood pressure to be equivalent to the calibration value itself. If our proposed model fails to outperform this basic model, it would indicate that our model is not truly learning and therefore not justifying its complexity.

Consequently, we plan to evaluate both our N-PPG-based and N-IABP-based models against this baseline model, defined as $P_{\text{inference}} = P_{\text{calibration}}$

\subsection{Evaluation}

The AAMI (Association for the Advancement of Medical Instrumentation) \cite{white1993national} and the BHS (British Hypertension Society) \cite{o1993british} standards serve as crucial benchmarks for evaluating BP measurement devices. Although these standards are not specifically designed for PPG-based devices requiring calibration, they are widely adopted in literature and offer valuable insights into the efficacy of such methodologies.

The AAMI standard mandates that for systolic (SBP) and diastolic blood pressure (DBP), the mean difference (MD) between the reference and the device measurements should not exceed $\pm$5 mmHg, with a standard deviation (STD) of no more than 8 mmHg. This requirement is to be satisfied across a study population of at least 85 participants.

Conversely, the BHS standard introduces a grading system based on the cumulative percentages of absolute differences between reference and device measurements within specific thresholds ($\leq$5 mmHg, $\leq$10 mmHg, and $\leq$15 mmHg). Devices are graded from A to D based on their adherence to these criteria, with grade D representing performance that is inferior to grade C, as elaborated in Table \ref{tab:bhs}.

To provide a more nuanced evaluation of device performance, we also calculate the mean absolute error (MAE) and the Pearson correlation coefficient ($\rho$). This approach affords a comprehensive assessment of both the accuracy and the degree of correlation between device and reference measurements.

\begin{table}[]
\centering
\caption{\centering Grading criteria for BHS }
\label{tab:bhs}
\begin{tabular}{cccc}
\hline
\multirow{2}{*}{\textbf{Grade}} & \multicolumn{3}{c}{\textbf{Cumulative error (\%)}}             \\ \cline{2-4} 
                                & $\leq$ 5mmHg & $\leq$10mmHg & $\leq$15mmHg \\ \hline
A                               & 60                 & 85                  & 95                  \\
B                               & 50                 & 75                  & 90                  \\
C                               & 40                 & 65                  & 85                  \\
D                               & \multicolumn{3}{c}{Worse than C}                               \\ \hline
\multicolumn{4}{c}{BHS: British   Hypertension Society}                                         
\end{tabular}
\end{table}

\section{Results}
\label{Results}

We present the outcomes of our study, which utilized N-IABP and N-PPG signals as inputs, evaluated against the AAMI standards, including Mean Absolute Error (MAE) and Pearson correlation coefficient ($\rho$), detailed in Table \ref{tab:result_aami}. Additionally, our evaluation extended to the BHS standard metrics, outlined in Table \ref{tab:result_bhs}. These comparisons also incorporate results from our baseline model, which predicts BP values by merely repeating the calibration values.

Considering the N-IABP signal, in Table \ref{tab:result_aami}, we observe that the raw version achieves STD, MAE, and $\rho$ of 5.78, 4.55, and 0.81, respectively, for DBP; and 6.33, 4.97, and 0.93 for SBP. Thus, the normalized raw version meets the AAMI standard for both SBP and DBP. Performance drops when employing the 0.5 Hz - 10 Hz band-pass filtered N-IABP: for DBP (STD: 6.35, MAE: 4.90, $\rho$: 0.77) and for SBP (STD: 9.85, MAE: 7.68, $\rho$: 0.83), where the AAMI standard is met for DBP but not for SBP. Diminishing the frequency range further to 0.5 Hz - 3.5 Hz results in an even greater drop in performance: for DBP (STD: 7.31, MAE: 5.65, $\rho$: 0.68) and for SBP (STD:11.72, MAE: 9.01, $\rho$: 0.76), with only DBP meeting the AAMI standard.

Analysis of Table \ref{tab:result_bhs} reveals that the raw N-IABP signal secures a grade of A for both SBP and DBP evaluations. Applying a 0.5 Hz - 10 Hz band-pass filter to the N-IABP signal maintains an A grade for DBP but decreases to C for SBP. Further narrowing the filter to 0.5 Hz - 3.5 Hz reduces grades to B for DBP and D for SBP.

For the PPG signals, Table \ref{tab:result_aami} shows negligible differences between the raw and the 0.5 - 10 Hz band-pass normalized versions; however, a notable performance decline occurs with the 0.5 - 3.5 Hz band-pass filtering. The raw N-PPG version achieved STD, MAE, and $\rho$ of 7.27, 5.68, and 0.69, respectively, for DBP, and 11.82, 9.20, and 0.75 for SBP. The 0.5 Hz - 10 Hz filtered signal mirrors these results closely for both DBP and SBP. Conversely, the 0.5 Hz - 3.5 Hz filtered N-PPG signal exhibits a significant drop in performance across all metrics.

According to the BHS standard in Table \ref{tab:result_bhs}, the raw and 0.5 Hz - 10 Hz filtered N-PPG signals achieve a grade B for DBP, whereas the 0.5 Hz - 3.5 Hz filtered version receives a grade C. For SBP, all filtering scenarios -- raw, 0.5 Hz - 10 Hz, and 0.5 Hz - 3.5 Hz -- result in a grade D.

Comparing N-IABP with N-PPG signals, the latter consistently underperforms across all frequency ranges and metrics, except for the mean difference (MD), which remains comparable.

Nevertheless, it is important to note that both N-IABP and N-PPG-based models surpassed the baseline model in every assessed metric except for ME, indicating that the models effectively learn the correlation between the signals and blood pressure beyond simply replicating calibration values.

\begin{table*}[]
\centering
\caption{Evaluation of BP estimation in the VitalDB dataset with regard to AAMI standard and MAE.}
\label{tab:result_aami}
\begin{tabular}{cccccc|cccc}
\hline
\multirow{2}{*}{Signal} & \multirow{2}{*}{Filter [Hz]} & \multicolumn{4}{c}{DBP}       & \multicolumn{4}{c}{SBP}  \\\cline{3-10} 
                        &                         & MAE   & MD    & STD   & $\rho$   & MAE   & MD    & STD   & $\rho$   \\ \hline
\multirow{3}{*}{IABP}    & -                       & 4.55  & -1.17 & 5.78  & 0.81  & 4.97  & -1.29 & 6.33  & 0.93  \\
                        & FC: {[}0.5, 10{]}       & 4.90  & -0.67 & 6.35  & 0.77  & 7.68  & -1.85 & 9.85  & 0.83  \\
                        & FC: {[}0.5, 3.5{]}      & 5.65  & -0.89 & 7.31  & 0.68  & 9.01  & -1.50 & 11.72 & 0.76  \\ \hline
\multirow{3}{*}{PPG}    & -                       & 5.68  & -0.89 & 7.27  & 0.69  & 9.20  & -1.49 & 11.82 & 0.75  \\
                        & FC: {[}0.5, 10{]}       & 5.65  & -1.07 & 7.20  & 0.69  & 9.34  & -1.73 & 11.95 & 0.75  \\
                        & FC: {[}0.5, 3.5{]}      & 6.33  & -1.21 & 8.04  & 0.59  & 10.92 & -1.40 & 14.04 & 0.63 \\ \hline
Baseline                & -                       & 8.39  & 0.00  & 11.08 & 0.36  & 16.23 & 0.00  & 20.75 & 0.31  \\ \hline
\multicolumn{10}{p{370pt}}{ AAMI: Association for the Advancement of Medical Instrumentation; DBP: Diastolic Blood Pressure; SBP: Systolic Blood Pressure; MD: Mean Difference;  STD: Standard Deviation; MAE: Mean Absolute Error; $\rho$: Pearson correlation coefficient }
\end{tabular}
\end{table*}

\begin{table*}[]
\centering
\caption{ Evaluation of BP estimation in the VitalDB dataset with regard to BHS standard.}
\label{tab:result_bhs}
\begin{tabular}{ccccc|ccc}
\hline
\multirow{2}{*}{Signal} & \multirow{2}{*}{Filter [Hz]} & \multicolumn{3}{c}{DBP}    & \multicolumn{3}{c}{SBP}   \\ \cline{3-8} 
                        &                         & $\leq$5 mmHg & $\leq$10 mmHg & $\leq$15 mmHg & $\leq$5 mmHg & $\leq$10 mmHg & $\leq$15 mmHg \\ \hline
\multirow{3}{*}{N-IABP}    & -                       & 63\%   & 92\%    & 98\%    & 60\%   & 89\%    & 97\%    \\
                        & FC: {[}0.5, 10{]}       & 60\%   & 89\%    & 97\%    & 42\%   & 72\%    & 88\%    \\
                        & FC: {[}0.5, 3.5{]}      & 54\%   & 84\%    & 95\%    & 37\%   & 65\%    & 82\%    \\
\multirow{3}{*}{N-PPG}    & -                       & 53\%   & 84\%    & 96\%    & 35\%   & 63\%    & 81\%    \\
                        & FC: {[}0.5, 10{]}       & 54\%   & 85\%    & 96\%    & 35\%   & 63\%    & 81\%    \\
                        & FC: {[}0.5, 3.5{]}      & 48\%   & 80\%    & 93\%    & 30\%   & 56\%    & 74\%   \\ \hline
Baseline                & -                       & 40\%   & 66\%    & 83\%    & 22\%   & 39\%    & 54\%    \\ \hline
\end{tabular}
\end{table*}

\section{Discussion}
\label{Discussion}
%
Our study sets out to delineate the potential and constraints of PPG signals in estimating blood pressure. 
We integrated max-min normalized signals (N-PPG and N-IABP) within a Siamese-based deep learning architecture to predict systolic and diastolic blood pressure through a calibration-based methodology. 
Given the significant correlation between PPG and IABP signals \cite{martinez2018}, which stems from their shared origin in cardiac contractions, we postulate that these signals likely relay similar morphological blood pressure information over time. 
Thus, if blood pressure prediction from N-IABP signals is found to be unfeasible, it would likely imply a similar outcome for PPG signals. Consequently, our study compares the performances of N-PPG against N-IABP signals.

Utilizing the IABP signal for blood pressure prediction might initially seem counterintuitive, as the IABP signal intrinsically contains blood pressure information. Nonetheless, by applying max-min normalization, we effectively isolate morphological details from explicit systolic and diastolic readings. In assessing the efficacy of N-IABP signals in blood pressure extrapolation, we aim to establish a performance benchmark for PPG-based methodologies, providing a clearer perspective on the upper limits of achievable results from signal morphology.

Our analysis processed both normalized PPG and IABP signals through three distinct modalities: i) raw, without filtering; ii) applying a 4th-order Chebyshev Type II band-pass filter within a 0.5 Hz - 10 Hz frequency range; and iii) employing heart rate-specific filtering at 0.5 Hz - 3.5 Hz (30 BPM - 210 BPM). These varied strategies allowed us to scrutinize the impact of filtering on model performance, especially in terms of information fidelity. The unfiltered state presents practical challenges due to noise interference, whereas the 0.5 Hz - 10 Hz filtered state reflects a common practice in existing literature \cite{li2018comparison, pollreisz2022detection}. The narrow 0.5 Hz - 3.5 Hz filtering aims to evaluate the model's reliance on heart rate information, countering the narrative that PPG-derived blood pressure insights are predominantly heart rate-centric \cite{mehta2023can, hasanzadeh2023hypertension}. Each filtered signal was then subjected to max-min normalization to ensure uniform comparison.

A baseline scenario, wherein predictions merely repeat a calibration value, was also evaluated. This benchmark is crucial to gauge the additive value of our model beyond mere calibration replication, affirming its applicability in real-world settings. Across all scenarios, our models outperformed this baseline, indicating a successful learning of correlations between signals and blood pressure values that extend beyond calibration data alone.

Performance degradation observed with narrowing filter ranges on N-IABP signals suggests the potential loss of critical blood pressure information, underscoring the need for PPG-specific filtering considerations. Although a 4th-order Chebyshev Type II filter is recommended \cite{liang2018optimal}, its application paradoxically diminished predictive accuracy, highlighting the delicate balance required in filter selection.

For N-PPG signals, minimal performance variance was noted between raw and 0.5 Hz - 10  Hz filtered versions, diverging from N-IABP signal outcomes. This discrepancy might derive from the VitalDB equipment's native handling of N-IABP signals, which foregoes filtering to preserve waveform integrity—and by extension, blood pressure accuracy. In contrast, PPG signals sourced from VitalDB are presumed to undergo inherent filtering, aligning our 'raw' PPG rendition with the 0.5 Hz - 10 Hz filtered profile. A significant performance drop with the 0.5 Hz - 3.5 Hz N-PPG filtering indicates that overly restrictive frequency range filtration adversely affects outcomes.

The finding that the 0.5 Hz - 3.5 Hz N-IABP signal still yields comparably robust results against all PPG signal frequencies challenges the notion that heart rate data alone encapsulates this frequency range's informational content. This observation suggests that additional, non-heart rate information is present, questioning initial assumptions. Moreover, the analogous performance of the 0.5 Hz - 3.5 Hz N-IABP and N-PPG signals (raw and 0.5-10Hz filtered) suggest that PPG signals may indeed represent a filtered manifestation of the IABP signal, as posited in prior studies \cite{zhang2018reconstruction, shi2023hybrid} , potentially missing crucial blood pressure-related data due to this inherent filtration.

In summary, our findings affirm presence of blood-pressure related information in PPG signals to some extent but are not sure they are sufficient for blood pressure estimation. The PPG signal did not achieve the AAMI standards and did not get a reasonable grade in the BHS standard, but it was able to add information beyond the calibration value for predicting blood pressure. The normalized N-IABP signals offer better results than the PPG signal, since it is more related to blood pressure than PPG signals, this provides a perspective on the best achievable results using signal morphology as input.
In this context, our work sets realistic expectations on the use of PPG signals for future research and application in consumer health technology.

However, our study's limitations must be acknowledged. We relied on a singular dataset from surgical patients, which may not accurately reflect the broader population's hemodynamics. These findings, therefore, should be contextualized within this specific patient group.

\section{Conclusion}
\label{Conclusion}

We successfully developed a calibration-based model for predicting blood pressure from input signals, specifically normalized PPG (N-PPG) and normalized IABP (N-IABP) signals. Our methodology employs a ResNet-based architecture within a Siamese-like framework. This approach processes a calibration signal alongside its corresponding systolic and diastolic pressures, as well as the target signal for inference.

To ensure the integrity of our validation process, we meticulously arranged our data to prevent the same patient's signals from appearing across different data splits. This strategy effectively mitigates the risk of data leakage and the potential for overoptimistic outcomes, as outlined in Dias et al. \cite{dias2022machine}.

Our findings indicate that our model outperforms the established baseline, demonstrating its capability to discern the intricate relationship between the input signals and the resulting blood pressure readings. Notably, the N-IABP signals yielded superior performance compared to N-PPG signals. This outcome aligns with expectations, considering that N-IABP signals encapsulate the blood pressure waveform's morphology absent of explicit systolic and diastolic values. The comparative success of N-IABP signals underscores the potential limitations and relative performance ceiling for N-PPG signals within this context.

Our results using filtered signals on different filter bands reveals that overly restrictive frequency bands can detrimentally impact blood pressure estimation accuracy for both N-PPG and N-IABP signals. This finding prompts a reevaluation of the optimal filter parameters for blood pressure estimation, challenging previous assertions by Liang et al. \cite{liang2018optimal} regarding the efficacy of specific filters for PPG signals.

The increase interest in PPG technology for integrating various biomarkers into consumer wearables is a testament to its potential. However, the nuanced performance of N-IABP signals in our study tempers expectations, highlighting the inherent challenges in blood pressure estimation even with direct waveform analysis.
As PPG technology continues to dominate discussions on non-invasive blood pressure monitoring, our study emphasizes the need for a critical assessment of its limitations. 

\section*{Authors contributions statement}
F.M.D., D.A.C.C. and M.A.F.T. Conceptualization, Methodology, Implementation and Writting. F.A.C.O. and E.R. Writting and Review. J.E.K. and M.A.G. Supervision and Review. All authors analyzed the results and revised critically the manuscript. All authors read and approved the submitted manuscript.

\section*{Competing interests}
The authors declare no competing interests.

\section*{Data availability}

\section*{Code availability}
Codes will be made available on request.

\section*{Acknowledgements}
This study supported by the Foxconn Brazil, and the Zerbini Foundation as part of the research project ``Machine Learning in Cardiovascular Medicine''.

\bibliographystyle{unsrt}
\bibliography{reference}  

\begin{thebibliography}{10}

\bibitem{CHOCKALINGAM2007}
Arun Chockalingam.
\newblock Impact of world hypertension day.
\newblock {\em Canadian Journal of Cardiology}, 23(7):517--519, 2007.

\bibitem{carey2022treatment}
Robert~M Carey, Andrew~E Moran, and Paul~K Whelton.
\newblock Treatment of hypertension: a review.
\newblock {\em Jama}, 328(18):1849--1861, 2022.

\bibitem{park2019ideal}
Sungha Park.
\newblock Ideal target blood pressure in hypertension.
\newblock {\em Korean circulation journal}, 49(11):1002--1009, 2019.

\bibitem{burton1967criterion}
AC~Burton.
\newblock The criterion for diastolic pressure-revolution and counterrevolution.
\newblock {\em Circulation}, 36(6):805--809, 1967.

\bibitem{o1990british}
Eoin O'Brien, James Petrie, William Littler, Michael de~Swiet, Paul~L Padfield, Kevin O'Malley, Michael Jamieson, Douglas Altman, Martin Bland, and Neil Atkins.
\newblock The british hypertension society protocol for the evaluation of automated and semi-automated blood pressure measuring devices with special reference to ambulatory systems.
\newblock {\em Journal of hypertension}, 8(7):607--619, 1990.

\bibitem{jones2003measuring}
Daniel~W Jones, Lawrence~J Appel, Sheldon~G Sheps, Edward~J Roccella, and Claude Lenfant.
\newblock Measuring blood pressure accurately: new and persistent challenges.
\newblock {\em Jama}, 289(8):1027--1030, 2003.

\bibitem{parati2006blood}
Gianfranco Parati, Andrea Faini, and Mariaconsuelo Valentini.
\newblock Blood pressure variability: its measurement and significance in hypertension.
\newblock {\em Current hypertension reports}, 8(3):199--204, 2006.

\bibitem{schroeder2010cardiovascular}
Rebecca~A Schroeder, Atilio Barbeito, Shahar Bar-Yosef, and Jonathan~B Mark.
\newblock Cardiovascular monitoring.
\newblock {\em Miller's anesthesia}, 7:1267--328, 2010.

\bibitem{scheer2002clinical}
Bernd~Volker Scheer, Azriel Perel, and Ulrich~J. Pfeiffer.
\newblock Clinical review: Complications and risk factors of peripheral arterial catheters used for haemodynamic monitoring in anaesthesia and intensive care medicine.
\newblock {\em Critical Care}, 6(199):198--204, 2002.

\bibitem{el2020review}
Chadi El-Hajj and Panayiotis~A Kyriacou.
\newblock A review of machine learning techniques in photoplethysmography for the non-invasive cuff-less measurement of blood pressure.
\newblock {\em Biomedical Signal Processing and Control}, 58:101870, 2020.

\bibitem{mukkamala2022cuffless}
Ramakrishna Mukkamala, George~S Stergiou, and Alberto~P Avolio.
\newblock Cuffless blood pressure measurement.
\newblock {\em Annual Review of Biomedical Engineering}, 24:203--230, 2022.

\bibitem{deb2007cuff}
Sujay Deb, Chinmayee Nanda, D~Goswami, J~Mukhopadhyay, and S~Chakrabarti.
\newblock Cuff-less estimation of blood pressure using pulse transit time and pre-ejection period.
\newblock In {\em 2007 International Conference on Convergence Information Technology (ICCIT 2007)}, pages 941--944. IEEE, 2007.

\bibitem{jc1922velocity}
BRAMWELL JC.
\newblock The velocity of the pulse wave in man.
\newblock {\em Proc. R. Soc. Lond (Biol)}, 93:298--306, 1922.

\bibitem{tijsseling2012isebree}
AS~Tijsseling and A~Anderson.
\newblock A. isebree moens and dj korteweg: on the speed of propagation of waves in elastic tubes.
\newblock 2012.

\bibitem{mousavi2019blood}
Seyedeh~Somayyeh Mousavi, Mohammad Firouzmand, Mostafa Charmi, Mohammad Hemmati, Maryam Moghadam, and Yadollah Ghorbani.
\newblock Blood pressure estimation from appropriate and inappropriate ppg signals using a whole-based method.
\newblock {\em Biomedical Signal Processing and Control}, 47:196--206, 2019.

\bibitem{hasanzadeh2019blood}
Navid Hasanzadeh, Mohammad~Mahdi Ahmadi, and Hoda Mohammadzade.
\newblock Blood pressure estimation using photoplethysmogram signal and its morphological features.
\newblock {\em IEEE Sensors Journal}, 20(8):4300--4310, 2019.

\bibitem{slapnivcar2019blood}
Ga{\v{s}}per Slapni{\v{c}}ar, Nejc Mlakar, and Mitja Lu{\v{s}}trek.
\newblock Blood pressure estimation from photoplethysmogram using a spectro-temporal deep neural network.
\newblock {\em Sensors}, 19(15):3420, 2019.

\bibitem{panwar2020pp}
Madhuri Panwar, Arvind Gautam, Dwaipayan Biswas, and Amit Acharyya.
\newblock Pp-net: A deep learning framework for ppg-based blood pressure and heart rate estimation.
\newblock {\em IEEE Sensors Journal}, 20(17):10000--10011, 2020.

\bibitem{aguet2021feature}
Clementine Aguet, Jerome Van~Zaen, Joao Jorge, Martin Proenca, Guillaume Bonnier, Pascal Frossard, and Mathieu Lemay.
\newblock Feature learning for blood pressure estimation from photoplethysmography.
\newblock In {\em 2021 43rd Annual International Conference of the IEEE Engineering in Medicine \& Biology Society (EMBC)}, pages 463--466. IEEE, 2021.

\bibitem{martinez2018}
Gloria Martínez, Newton Howard, Derek Abbott, Kenneth Lim, Rabab Ward, and Mohamed Elgendi.
\newblock Can photoplethysmography replace arterial blood pressure in the assessment of blood pressure?
\newblock {\em Journal of Clinical Medicine}, 7(10), 2018.

\bibitem{ibtehaz2020ppg2ABP}
Nabil Ibtehaz, Sakib Mahmud, Muhammad~EH Chowdhury, Amith Khandakar, Mohamed~Arselene Ayari, Anas Tahir, and M~Sohel Rahman.
\newblock Ppg2abp: Translating photoplethysmogram (ppg) signals to arterial blood pressure (abp) waveforms using fully convolutional neural networks.
\newblock {\em arXiv preprint arXiv:2005.01669}, 2020.

\bibitem{dias2022machine}
Felipe~M Dias, Thiago~BS Costa, Diego~AC Cardenas, Marcelo~AF Toledo, Jose~E Krieger, and Marco~A Gutierrez.
\newblock A machine learning approach to predict arterial blood pressure from photoplethysmography signal.
\newblock In {\em 2022 Computing in Cardiology (CinC)}, volume 498, pages 1--4. IEEE, 2022.

\bibitem{da2023blood}
Thiago~Bulh{\~o}es da~Silva~Costa, Felipe~Meneguitti Dias, Diego Armando~Cardona Cardenas, Marcelo Arruda~Fiuza De~Toledo, Daniel~M{\'a}rio De~Lima, Jose~Eduardo Krieger, and Marco~Antonio Gutierrez.
\newblock Blood pressure estimation from photoplethysmography by considering intra-and inter-subject variabilities: guidelines for a fair assessment.
\newblock {\em IEEE Access}, 2023.

\bibitem{mehta2023can}
Suril Mehta, Nipun Kwatra, Mohit Jain, and Daniel McDuff.
\newblock " can't take the pressure?": Examining the challenges of blood pressure estimation via pulse wave analysis.
\newblock {\em arXiv preprint arXiv:2304.14916}, 2023.

\bibitem{weber2023intensive}
Guillaume Weber-Boisvert, Benoit Gosselin, and Frida Sandberg.
\newblock Intensive care photoplethysmogram datasets and machine-learning for blood pressure estimation: Generalization not guarantied.
\newblock {\em Frontiers in Physiology}, 14:317, 2023.

\bibitem{lee2022vitaldb}
Hyung-Chul Lee, Yoonsang Park, Soo~Bin Yoon, Seong~Mi Yang, Dongnyeok Park, and Chul-Woo Jung.
\newblock Vitaldb, a high-fidelity multi-parameter vital signs database in surgical patients.
\newblock {\em Scientific Data}, 9(1):279, 2022.

\bibitem{kachuee2015cuff}
Mohamad Kachuee, Mohammad~Mahdi Kiani, Hoda Mohammadzade, and Mahdi Shabany.
\newblock Cuff-less high-accuracy calibration-free blood pressure estimation using pulse transit time.
\newblock In {\em 2015 IEEE international symposium on circuits and systems (ISCAS)}, pages 1006--1009. IEEE, 2015.

\bibitem{kachuee2016cuffless}
Mohammad Kachuee, Mohammad~Mahdi Kiani, Hoda Mohammadzade, and Mahdi Shabany.
\newblock Cuffless blood pressure estimation algorithms for continuous health-care monitoring.
\newblock {\em IEEE Transactions on Biomedical Engineering}, 64(4):859--869, 2016.

\bibitem{cuff}
Mohamad Kachuee, Mohammad Kiani, Hoda Mohammadzade, and Mahdi Shabany.
\newblock {Cuff-Less Blood Pressure Estimation}.
\newblock \url{UCI Machine Learning Repository}, 2015.
\newblock \url{https://doi.org/10.24432/C5B602}.

\bibitem{wang2024imsf}
Di~Wang, Yutong Ye, Bowen Zhang, Jinlu Sun, and Cheng Zhang.
\newblock Imsf-net: An improved multi-scale information fusion network for ppg-based blood pressure estimation.
\newblock {\em Biomedical Signal Processing and Control}, 90:105791, 2024.

\bibitem{el2021cuffless}
Chadi El-Hajj and Panayiotis~A Kyriacou.
\newblock Cuffless blood pressure estimation from ppg signals and its derivatives using deep learning models.
\newblock {\em Biomedical Signal Processing and Control}, 70:102984, 2021.

\bibitem{saeed2011multiparameter}
Mohammed Saeed, Mauricio Villarroel, Andrew~T Reisner, Gari Clifford, Li-Wei Lehman, George Moody, Thomas Heldt, Tin~H Kyaw, Benjamin Moody, and Roger~G Mark.
\newblock Multiparameter intelligent monitoring in intensive care ii (mimic-ii): a public-access intensive care unit database.
\newblock {\em Critical care medicine}, 39(5):952, 2011.

\bibitem{rolfe1979photoelectric}
P~Rolfe.
\newblock Photoelectric plethysmography for estimating cutaneous blood flow.
\newblock {\em New York: Academic}, pages 125--151, 1979.

\bibitem{CHARLTON2022401}
Peter~H. Charlton and Vaidotas Marozas.
\newblock 12 - wearable photoplethysmography devices.
\newblock In John Allen and Panicos Kyriacou, editors, {\em Photoplethysmography}, pages 401--439. Academic Press, 2022.

\bibitem{allen2007photoplethysmography}
John Allen.
\newblock Photoplethysmography and its application in clinical physiological measurement.
\newblock {\em Physiological measurement}, 28(3):R1, 2007.

\bibitem{park2022photoplethysmogram}
Junyung Park, Hyeon~Seok Seok, Sang-Su Kim, and Hangsik Shin.
\newblock Photoplethysmogram analysis and applications: An integrative review.
\newblock {\em Frontiers in Physiology}, 12:808451, 2022.

\bibitem{hartmann2019quantitative}
Vera Hartmann, Haipeng Liu, Fei Chen, Qian Qiu, Stephen Hughes, and Dingchang Zheng.
\newblock Quantitative comparison of photoplethysmographic waveform characteristics: Effect of measurement site.
\newblock {\em Frontiers in physiology}, 10:198, 2019.

\bibitem{AHA2021}
{American Heart Association}.
\newblock What is high blood pressure?
\newblock \url{https://www.heart.org/}, 2021.
\newblock Accessed: [Insert access date here].

\bibitem{meidert2018techniques}
Agnes~S Meidert and Bernd Saugel.
\newblock Techniques for non-invasive monitoring of arterial blood pressure.
\newblock {\em Frontiers in medicine}, 4:231, 2018.

\bibitem{goodman2023measuring}
Chris~TD Goodman and Gareth~B Kitchen.
\newblock Measuring arterial blood pressure.
\newblock {\em Anaesthesia \& Intensive Care Medicine}, 2023.

\bibitem{lakhal2012noninvasive}
Karim Lakhal, Christine Macq, Stephan Ehrmann, Thierry Boulain, and Xavier Capdevila.
\newblock Noninvasive monitoring of blood pressure in the critically ill: reliability according to the cuff site (arm, thigh, or ankle).
\newblock {\em Critical care medicine}, 40(4):1207--1213, 2012.

\bibitem{WARD2007122}
Matthew Ward and Jeremy~A Langton.
\newblock Blood pressure measurement.
\newblock {\em Continuing Education in Anaesthesia Critical Care \& Pain}, 7(4):122--126, 2007.

\bibitem{bishop2018multi}
Steven~M Bishop and Ari Ercole.
\newblock Multi-scale peak and trough detection optimised for periodic and quasi-periodic neuroscience data.
\newblock In {\em Intracranial Pressure \& Neuromonitoring XVI}, pages 189--195. Springer, 2018.

\bibitem{liang2018optimal}
Yongbo Liang, Mohamed Elgendi, Zhencheng Chen, and Rabab Ward.
\newblock An optimal filter for short photoplethysmogram signals.
\newblock {\em Scientific data}, 5(1):1--12, 2018.

\bibitem{ribeiro2020automatic}
Ant{\^o}nio~H Ribeiro, Manoel~Horta Ribeiro, Gabriela~MM Paix{\~a}o, Derick~M Oliveira, Paulo~R Gomes, J{\'e}ssica~A Canazart, Milton~PS Ferreira, Carl~R Andersson, Peter~W Macfarlane, Wagner Meira~Jr, et~al.
\newblock Automatic diagnosis of the 12-lead ecg using a deep neural network.
\newblock {\em Nature communications}, 11(1):1760, 2020.

\bibitem{white1993national}
William~B White, Alan~S Berson, Carroll Robbins, Michael~J Jamieson, L~Michael Prisant, Edward Roccella, and Sheldon~G Sheps.
\newblock National standard for measurement of resting and ambulatory blood pressures with automated sphygmomanometers.
\newblock {\em Hypertension}, 21(4):504--509, 1993.

\bibitem{o1993british}
Eoin O’Brien, James Petrie, William Littler, Michael De~Swiet, Paul~L Padfield, Douglas Altman, Martin Bland, Andrew Coats, Neil Atkins, et~al.
\newblock The british hypertension society protocol for the evaluation of blood pressure measuring devices.
\newblock {\em J hypertens}, 11(Suppl 2):S43--S62, 1993.

\bibitem{li2018comparison}
Suyi Li, Lijia Liu, Jiang Wu, Bingyi Tang, and Dongsheng Li.
\newblock Comparison and noise suppression of the transmitted and reflected photoplethysmography signals.
\newblock {\em BioMed research international}, 2018, 2018.

\bibitem{pollreisz2022detection}
David Pollreisz and Nima TaheriNejad.
\newblock Detection and removal of motion artifacts in ppg signals.
\newblock {\em Mobile Networks and Applications}, 27(2):728--738, 2022.

\bibitem{hasanzadeh2023hypertension}
Navid Hasanzadeh, Shahrokh Valaee, and Hojjat Salehinejad.
\newblock Hypertension detection from high-dimensional representation of photoplethysmogram signals.
\newblock In {\em 2023 IEEE EMBS International Conference on Biomedical and Health Informatics (BHI)}, pages 1--4. IEEE, 2023.

\bibitem{zhang2018reconstruction}
Pandeng Zhang, Quanli Qiu, and Yanxia Zhou.
\newblock Reconstruction of continuous brachial artery pressure wave from continuous finger arterial pressure in humans.
\newblock {\em Australasian Physical \& Engineering Sciences in Medicine}, 41(4):1115--1125, 2018.

\bibitem{shi2023hybrid}
Wenying Shi, Congcong Zhou, Yiming Zhang, Kaitai Li, Xianglin Ren, Hui Liu, and Xuesong Ye.
\newblock Hybrid modeling on reconstitution of continuous arterial blood pressure using finger photoplethysmography.
\newblock {\em Biomedical Signal Processing and Control}, 85:104972, 2023.

\end{thebibliography}






\end{document}